\documentclass[a4paper,fleqn]{cas-dc}
\usepackage[square,sort&compress,comma,numbers
]{natbib}

\usepackage{bm}% bold math
\usepackage{amsmath}
\usepackage{amssymb}
\usepackage{amsfonts}
\usepackage{yfonts}
\usepackage{slashed}
\usepackage{nicefrac}

\newcommand{\bea}{\begin{eqnarray}}
\newcommand{\eea}{\end{eqnarray}}
\newcommand{\bel}[1]{\begin{eqnarray}\label{#1}}
\newcommand{\eel}{\end{eqnarray}}

\def\LB{\left(}
\def\RB{\right)}
\def\LSB{\left[}
\def\RSB{\right]}

\newcommand{\EQ}[1]{Eq.~(\ref{#1})}
\newcommand{\EQn}[1]{(\ref{#1})}

\newcommand{\EQSM}[2]{Eqs.~(\ref{#1})--(\ref{#2})}
\newcommand{\EQSMn}[2]{(\ref{#1})--(\ref{#2})}

\newcommand{\CIT}[1]{Ref.~\citep{#1}} 
\newcommand{\CITn}[1]{\citep{#1}} 

\newcommand{\p}{\partial}
\newcommand{\dd}{\mathrm{d}}

\newcommand{\bv}{{\boldsymbol b}} 
\newcommand{\bvp}{{\boldsymbol b}^\prime} 
\newcommand{\ev}{{\boldsymbol e}}
\newcommand{\evp}{{\boldsymbol e}^\prime}

\newcommand{\uv}{{\boldsymbol u}}
\newcommand{\vv}{{\boldsymbol v}}
\newcommand{\pv}{{\boldsymbol p}}
\newcommand{\pvp}{{\boldsymbol p}^\prime}

\newcommand{\f}[2]{\frac{#1}{#2}}
\newcommand{\onehalf}{{\nicefrac{1}{2}}}

\def\spin{\,\textgoth{s:}}

\def\ubarrp{{\bar u}_r(p)}
\def\ubarsp{{\bar u}_s(p)}
\def\usp{u_s(p)}
\def\urp{u_r(p)}

\def\vbarrp{{\bar v}_r(p)}
\def\vbarsp{{\bar v}_s(p)}
\def\vsp{v_s(p)}
\def\vrp{v_r(p)}

\newcommand{\trf}{{\rm tr_4}}

\begin{document}
\let\WriteBookmarks\relax
\def\floatpagepagefraction{1}
\def\textpagefraction{.001}

\shorttitle{Spin hydrodynamics -- recent developments}    
\shortauthors{S.~Bhadury, Z.~Drogosz, F.~Florkowski, V.~Mykhaylova}  
\title [mode = title]{Spin hydrodynamics -- recent developments}  
\author[uj]{Samapan Bhadury}[orcid=0000-0002-8693-4482]
\ead{samapan.bhadury@uj.edu.pl}

\author[uj]{Zbigniew Drogosz}[orcid=0000-0001-5133-958X]
\ead{zbigniew.drogosz@alumni.uj.edu.pl}

\author[uj]{Wojciech Florkowski}[orcid=0000-0002-9215-0238]
\ead{wojciech.florkowski@uj.edu.pl}

\author[uj]{Valeriya Mykhaylova}[orcid=0000-0002-2660-6954]
\ead{valeriya.mykhaylova@uj.edu.pl}

\affiliation[uj]{organization={Institute of Theoretical Physics, Jagiellonian University},
            %addressline={}, 
            city={Kraków},
            postcode={30-348}, 
            country={Poland}}

\begin{abstract}
After briefly touching on relativistic hydrodynamics, we provide a detailed description of recent developments in spin hydrodynamics. We discuss the theory of perfect spin hydrodynamics within two different approaches, which lead to identical generalized thermodynamic relations. We also indicate the applicability range of the theory, finding it compatible with the
conditions existing in the late stages of heavy-ion collisions. Finally, we discuss the near-equilibrium dynamics.

\end{abstract}

\maketitle

%%%%%%%%%%%%%%%%%%%%%%%%% main text
\section{Introduction}
\label{sec:intro}

Over the last two decades, the analysis of relativistic heavy-ion collisions in experimental facilities such as RHIC at BNL and the LHC at CERN has helped us to understand the properties of strongly interacting matter consisting of quarks and gluons. Directly probing this matter is a very difficult task because of the extremely short length and time scales during which the studied system evolves. 

Since experimental observables involve mainly final-state particles, one has to perform simulations of strongly interacting nonequilibrium matter through various evolution stages~\cite{Busza:2018rrf}. Relativistic hydrodynamics has become an indispensable tool in this regard \cite{Romatschke:2007mq, Romatschke:2009im, Jaiswal:2016hex, Florkowski:2017olj}. Consequently, numerous efforts have been made in recent years with the aim of developing the very formalism of relativistic hydrodynamics. Although the first formulations of relativistic hydrodynamics by Landau \cite{landau1987fluid} and later by Eckart \cite{Eckart:1940te} were constructed during the 1950s, it has turned out that these early frameworks require substantial modifications. 

The first formulations of relativistic dissipative hydrodynamics were a straightforward generalization of the non-relativistic Navier-Stokes theories. They were plagued by the problems of acausality \cite{Hiscock:1983zz, Hiscock:1985zz} driven by the parabolic nature of the equations of motion. Later, second-order theories of relativistic hydrodynamics were developed by M{\"u}ller, Israel, and Stewart (collectively known as the MIS theories) by promoting dissipative currents to new degrees of freedom of hydrodynamics \cite{Muller:1967zza, Israel:1976tn, Israel:1979wp}. 

In the advent of heavy-ion collision experiments, renewed interest in relativistic hydrodynamics has led to new theories such as BRSSS~\cite{Baier:2007ix}, DNMR~\cite{Denicol:2012cn}, anisotropic hydrodynamics~\cite{Florkowski:2010cf, Martinez:2010sc}, higher-order theories~\cite{El:2009vj, Jaiswal:2013vta}, and several other frameworks. Recently, in a series of papers \cite{Bemfica:2017wps, Kovtun:2019hdm, Bemfica:2019knx}, first-order causal and stable theories have been proposed, known as the BDNK approach. They do not introduce new degrees of freedom as compared to standard hydrodynamics. However, all of these theories generate non-hydrodynamic modes.

Interestingly, a new set of theories of hydrodynamics has been developed that contain only hydrodynamic modes and are suitable for the description of fluctuations in a hydrodynamic setup~\cite{Armas:2020mpr, Basar:2024qxd, Bhattacharyya:2024ohn}. However, these theories are not manifestly covariant. Several other formulations of relativistic hydrodynamics have been developed to describe specific systems. They include, in particular, magnetohydrodynamics \cite{Denicol:2018rbw, Panda:2020zhr}, hydro+ theories \cite{Stephanov:2017ghc, An:2019osr}, and spin hydrodynamics that is discussed in greater detail below.

Hydrodynamics has been very successful in describing the evolution of systems in which the microscopic scales (e.g., mean free path) are well separated from the macroscopic scales (e.g., space extension) of the system. A~more modern understanding of the theory is that hydrodynamics becomes applicable if all the fast degrees of freedom decay and only the slow degrees of freedom associated with the conserved currents survive. This is a consequence of the global symmetries of the system. 

In the context of relativistic heavy-ion collision experiments, the most relevant symmetries include: time and space translations, Lorentz transformations (boosts and rotations), and $U(1)$ symmetries connected with conserved charges $Q$ (electric charge, baryon number, and strangeness). These symmetries lead to the conservation of the energy-momentum tensor ($T^{\mu\nu}$), the total angular momentum tensor ($J^{\lambda,\mu\nu}$), and the charge currents ($N_Q^\mu$), respectively. In the following, we consider only the conservation of the baryon number current identified with $N_B^\mu = N^\mu$.

For scalar particles, which have no internal angular momentum, the angular momentum conservation law is redundant (as it follows from the energy-momentum conservation). However, for particles with finite spin, the total angular momentum can be split into an orbital and a spin part. This case gives an opportunity to develop a spin hydrodynamic theory, where the conservation law $\partial_\lambda J^{\lambda,\mu\nu} = 0$ is nontrivial to fulfill and is treated on the same footing as the conservation law $\partial_\mu T^{\mu\nu} = 0$ used in traditional hydrodynamics. 

Interestingly, prompted by the recent discovery of the spin-polarization phenomena at RHIC \cite{STAR:2017ckg, Adam:2018ivw, Niida:2018hfw}, spin hydrodynamics has received a fair amount of attention. However, unlike the traditional theory of relativistic hydrodynamics, whose various formulations are well-established~\cite{Florkowski:2017olj,Romatschke:2017ejr}, the spin hydrodynamics is being developed along different paths, mutual relations between which are not completely understood at the moment. The most popular current frameworks are: {\bf (i)} the approach employing gradients of the standard hydrodynamic variables characterizing the spin polarization~\CITn{Becattini:2013fla,  Becattini:2021iol, Palermo:2024tza}. The main quantities used are thermal vorticity
\begin{equation}\label{thvort}
\varpi_{\mu \nu} = -\frac{1}{2} (\p_\mu \beta_\nu-\p_\nu \beta_\mu)   
\end{equation}
and thermal shear 
\begin{equation}\label{thsh}
\xi_{\mu \nu} = -\frac{1}{2} (\p_\mu \beta_\nu+\p_\nu \beta_\mu),
\end{equation}
where $\beta_\mu = u_\mu/T$, with $u_\mu$ and $T$ being the hydrodynamic flow and temperature. {\bf (ii)} Kinetic theory, from which the spin hydrodynamic equations are derived~\mbox{\CITn{Florkowski:2017ruc, Florkowski:2017dyn, Florkowski:2018ahw, Florkowski:2018fap, Bhadury:2020puc, Bhadury:2022ulr, Weickgenannt:2019dks, Weickgenannt:2021cuo, Weickgenannt:2020aaf, Weickgenannt:2022zxs, Weickgenannt:2023nge, Wagner:2024fhf, Hu:2021pwh, Li:2020eon, Shi:2020htn, Banerjee:2024xnd, Bhadury:2024ckc}.} {\bf (iii)}~The~formalism developed by making reference to the mathematically allowed forms for the energy-momentum and spin tensors, following the Israel-Stewart \sloppy method~\mbox{\CITn{Hattori:2019lfp, Fukushima:2020ucl, Daher:2022xon, Daher:2022wzf, Sarwar:2022yzs, Wang:2021ngp, Biswas:2023qsw, Xie:2023gbo, Daher:2024ixz, Ren:2024pur, Daher:2024bah, Gallegos:2021bzp, Hongo:2021ona, Kumar:2023ojl, She:2021lhe}}. {\bf (iv)} A spin-extended Lagrangian formalism~\CITn{Montenegro:2017rbu}. For a recent review, see, for exmaple, \CIT{Huang:2024ffg}.

Various approaches to spin hydrodynamics result in different opinions on several important concepts. They include, in particular, the idea of local equilibrium for particles with spin, the method of hydrodynamic expansion, and the choice of the specific forms of the tensors used in spin hydrodynamics (the so-called pseudogauge dependence). In the present contribution, we give a brief overview of recent developments in spin hydrodynamics and discuss attempts to achieve convergence between its different formulations~\mbox{\cite{Florkowski:2024bfw, Drogosz:2024gzv, Florkowski:2024cif, Bhadury:2025boe, Drogosz:2025ihp}}. We note that herein we restrict ourselves to the description of systems consisting of spin-$1/2$ particles only, as the main objective of spin hydrodynamics is to explain the spin polarization of $\Lambda$ hyperons. 

\textit{Notations and conventions}: We work with natural units, $\hbar = c = k_B = 1$. We use the mostly negative metric convention $g^{\mu\nu} = {\rm diag} (1,-1,-1,-1)$ and the Levi-Civita symbol values $\varepsilon^{0123} = - \varepsilon_{0123} = 1$.

%%%%%%%%%%%%%%%%%%%%%%%%% main text
\section{Perfect spin hydrodynamics}
\label{sec:perf}

Similarly to standard perfect fluid hydrodynamics, we assume that perfect spin hydrodynamics is based on the conservation of baryon number, energy, linear momentum, and entropy. Moreover, following our earlier works~\mbox{\CITn{Florkowski:2017ruc,Florkowski:2018fap}}, we additionally assume that perfect spin hydrodynamics is based on the conservation of the spin part of the total angular momentum. In physics terms, this corresponds to a regime in which the interaction is dominated by the $s$-wave scattering $(l=0)$~\CITn{Coleman:2018mew}. Alternatively, one may think of systems where spin-changing processes are slow~\CITn{Kapusta:2019sad, Kapusta:2020npk, Banerjee:2024xnd}. 

Conservation of the spin tensor allows the introduction of the tensor spin chemical potential $\Omega_{\mu\nu}$, which can be interpreted as a tensor Lagrange multiplier (with \mbox{$\Omega_{\mu\nu}=-\Omega_{\nu\mu}$}). As a consequence, one can construct a framework of perfect spin hydrodynamics based on eleven conservation laws for the baryon number, energy, linear momentum, and spin part of the angular momentum,
\bel{eq:psh}
\p_\mu N^\mu_{\rm eq}  = 0, \quad \p_\mu T^{\mu \alpha}_{\rm eq} = 0, \quad \p_\mu S^{\mu,\alpha \beta}_{\rm eq} = 0,
\eel
with $N^\mu_{\rm eq}$, $T^{\mu \alpha}_{\rm eq}$, and $S^{\mu,\alpha \beta}_{\rm eq}$ denoting the baryon current, the energy-momentum tensor and the {\it spin tensor} in local equilibrium. The unknown functions to be determined by the hydrodynamic equations are: the baryon chemical potential $\mu$, temperature $T$, three independent components of the flow vector $\uv$, and the spin chemical potential $\Omega_{\mu\nu}$.

%%%%%%%%%%%%%%%%%%%%%%%%% main text
\section{Local equilibrium}
\label{sec:loceq}

Within kinetic theory, the framework of perfect spin hydrodynamics follows from the definition of the local equilibrium function. Two approaches have been explored.

%%%%%%%%%%%%%%%%%%%%%%%%% main text
\subsection{Classical spin treatment}
\label{sec:class_loceq}

The first one is based on the classical description of spin and uses the distribution function in an extended phase space,
\bel{eq:classEQ}
f^\pm_{\text{eq}}(x,p,s) = \exp\left[\pm \xi - \beta_\mu p^\mu + \frac{1}{2} \omega_{\alpha \beta} s^{\alpha \beta} \right],
\eel
where $\xi=\mu/T$, and $\pm$ refers to particles and antiparticles, respectively. The tensor $ s^{\alpha \beta} = (1/m) \epsilon^{\alpha\beta\gamma\delta} p_\gamma s_\delta$ describes internal angular momentum, with the spin vector $s_\delta$ satisfying the orthogonality condition $p_\mu s^\mu = 0$. The tensor $\omega_{\alpha \beta} = \Omega_{\alpha \beta}/T$ is called the {\it spin polarization tensor}. It plays a similar role to the ratio $\xi$ -- in natural units both $\xi$ and $\omega$ are dimensionless, which allows for expansions in $\xi$ and/or $\omega_{\alpha \beta}$. 

By computing moments of the distribution functions $f^\pm_{\text{eq}}(x,p,s)$, we obtain the macroscopic tensors, such as the baryon current
\bel{eq:Nmu} 
N_{\text{eq}}^\mu(x) = \int \dd P \,\dd S  \, p^\mu \left[f^+_{\text{eq}}(x,p,s)-f^-_{\text{eq}}(x,p,s) \right],
\eel
the energy-momentum tensor
\bel{eq:Tmunu} 
T_{\text{eq}}^{\mu\nu}(x) = \int \dd P \,\,\dd S \, p^\mu p^\nu \left[ f^+_{\text{eq}}(x,p,s)+f^-_{\text{eq}}(x,p,s) \right],
\eel
and the spin tensor
\bel{eq:Slmunu} 
S_{\text{eq}}^{\lambda, \mu\nu}(x) =  \int \dd P \,\dd S \, p^\lambda s^{\mu\nu} \left[ f^+_{\text{eq}}(x,p,s)  +  f^-_{\text{eq}}(x,p,s) \right].
\eel
Additionally, we define the particle current
\bel{eq:calNmu} 
{\cal N}_{\text{eq}}^\mu(x) =  \int \dd P \,\dd S  \, p^\mu \left[f^+_{\text{eq}}(x,p,s)  +  f^-_{\text{eq}}(x,p,s) \right],
\eel
and, following the original idea of Boltzmann~\CITn{landau1987fluid}, the entropy current
\bel{eq:Hmu1}
\hspace{-0.0cm} S_{\text{eq}}^\mu \!=\!-\!\!\int \!\!\dd P \, \dd S \, p^\mu 
\LSB 
f_{\text{eq}}^+ \LB \ln f_{\text{eq}}^+\!\!-\!1\RB\!+\! 
f_{\text{eq}}^- \LB \ln f_{\text{eq}}^-\!\!-1\!\RB \RSB.
\eel
In the spinless case one finds ${\cal N}^\mu = P \beta^\mu$, where $P$ is the pressure. In the expressions above, the following integration measures are used~\CITn{Florkowski:2018fap}
\bel{eq:dP_dS}
\hspace{-0.5cm} \dd P = \f{\dd^3p}{(2\pi)^3 E_{p}}, \quad  \dd S = \f{m}{\pi \spin}  \, \dd^4 s \, \delta(s \cdot s + \spin^2) \, \delta(p \cdot s),
\eel
where $E_{p}=\sqrt{\pv^2+m^2}$ is the energy of a particle of mass~$m$, and $\spin = \sqrt{3/4}$ is the eigenvalue of the SU(2) Casimir operator. We note that both $\dd P$ and $\dd S$ are Lorentz invariant.

%%%%%%%%%%%%%%%%%%%%%%%%% 
\subsection{Spin density approach}
\label{sec:qaunt_loceq}

The second approach uses the newly constructed local equilibrium Wigner function~\CITn{deGroot:1980dk, Bhadury:2025boe}
\bel{eq:Wpm}
\hspace{-0.5cm} W^\pm(x,k)\!=\!\frac{1}{4 m}  \int \!\dd P\,
\delta^{(4)}(k\!\mp\!p) 
(\slashed{p}\!\pm\! m) X^\pm
(\slashed{p}\!\pm\! m),
\eel
where
\bel{XpmNEW3}
X^{\pm}(x,p) =  \exp\left[\pm \xi(x) - \beta_\mu(x) p^\mu + \gamma_5 \slashed{a} \right]
\eel
is the spinor four-by-four density matrix with
\bel{eq:amu}
a_\mu(x,p) = -\frac{1}{2 m} {\tilde \omega}_{\mu\nu}(x)p^\nu.
\eel
Here ${\tilde \omega}_{\mu\nu}(x)$ is the dual polarization tensor. The orthogonality condition $a_\mu p^\mu = 0$ may be treated as a version of the Frenkel condition that is frequently used in spin hydrodynamics with different formulations~\CITn{Frenkel:1926zz}. 
Since $[\gamma_5 \slashed{a}, \slashed{p}] =0$, we can also rewrite~\EQ{eq:Wpm} as
\bel{eq:Wpm2} 
\hspace{-0.5cm} W^\pm(x,k) = \pm \frac{1}{2}  \int \dd P\,
\delta^{(4)}(k \mp p) (\slashed{p} \pm m) X^\pm.
\eel
It is interesting to note that the expressions $ (\slashed{p} \pm m) X_s^\pm$ are consistent with the expression for the spinor density matrix derived in the Landau-Lifshitz course~\CITn{Berestetskii:1982qgu}, as shown in~\CITn{Florkowski:2019gio}.

In the following it is convenient to introduce two-by-two spin density matrices
\bel{fplusrsxp}
f^+_{rs}(x,p) &=&  \frac{1}{2m} \,\ubarrp X^+ \usp,  \\
f^-_{rs}(x,p) &=& - \frac{1}{2m} \, \vbarsp X^- \vrp,
\label{fminusrsxp} 
\eel
where $\urp$ and $\vsp$ are Dirac spinors~\cite{Itzykson:1980rh}, while the bar denotes their Dirac adjoints. These matrices can be used to  derive the following expressions for the baryon current, the energy-momentum tensor, and the spin tensor~\CITn{deGroot:1980dk}:
\begin{align}
N_{\text{eq}}^\mu(x) &= \sum_{r=1}^2 \int \dd P  \, p^\mu \left[f^+_{rr}(x,p)-f^-_{rr}(x,p) \right],\\
T_{\text{eq}}^{\mu\nu}(x) &= \sum_{r=1}^2 \int \dd P \, p^\mu p^\nu \left[f^+_{rr}(x,p)+f^-_{rr}(x,p) \right],\\
\begin{split}
S_{\text{eq}}^{\lambda, \mu\nu}(x) &= \f{1}{2}\sum_{r,s=1}^2 \int \dd P \, p^\lambda  \left[\sigma^{+ \mu\nu}_{sr}(p) f^+_{rs}(x,p) \right. \\
 & \hspace{1.5cm} \left.   + \, \sigma^{- \mu\nu}_{sr} (p) f^-_{rs}(x,p) \right],  
\end{split}\end{align}
where~\cite{deGroot:1980dk}
\begin{equation}
\sigma^{+ \mu\nu}_{sr} (p) =\frac{1}{2m} \ubarsp \sigma^{\mu\nu} \urp    
\end{equation}
and
\begin{equation}
\sigma^{- \mu\nu}_{sr} (p) = \frac{1}{2m} \vbarrp \sigma^{\mu\nu} \vsp, 
\end{equation}
with 
\begin{equation}
\sigma^{\mu\nu}=\frac{i}{2}[\gamma^\mu,\gamma^\nu]. 
\end{equation}
We further define the particle current
\bel{eq:calNmu2} 
{\cal N}^\mu(x) = \sum_{r=1}^2 \int \dd P  \, p^\mu \left[f^+_{rr}(x,p) + f^-_{rr}(x,p) \right],
\eel
and the entropy current of the form derived in~\CIT{Florkowski:2017ruc},
\begin{align}\begin{split}\label{eq:Smu} 
S^\mu(x) &= -\f{1}{2} \int \dd P  \, p^\mu \left\{
\trf \, \left[ X^+  \left(\ln X^+ - 1\right) \right]\right. \\ 
& \left.  \hspace{1cm} + \, 
\trf \, \left[X^-  \left(\ln X^- - 1\right) \right]
\right\}.
\end{split}\end{align}

In the spinless case, we find ${\cal N}^\mu = P \beta^\mu$, as in Sec.~\ref{sec:class_loceq}. We note that because of the pseudogauge ambiguity, several different forms of the conserved tensors are used in the literature. The~forms introduced above use the GLW pseudogauge~\CITn{deGroot:1980dk}. 

%%%%%%%%%%%%%%%%%%%%%%%%% main text
\section{Generalized thermodynamic relations}
\label{sec:genrel}

It has been demonstrated in the recent works~\cite{Florkowski:2024bfw, Drogosz:2024gzv, Bhadury:2025boe} that thermodynamic relations used in spin local equilibrium should have a tensor form:
\bel{eq:Smu_FD1}
S^\mu_{\rm eq} = -N^\mu_{\rm eq} \xi + T^{\mu \alpha}_{\rm eq}\beta_\alpha - \frac{1}{2} S^{\mu,\alpha \beta}_{\rm eq}\omega_{\alpha \beta} + \mathcal{N}^\mu,
\eel
\bel{eq:dSmu}
\dd S^{\mu}_{\rm eq}= -\xi \dd N^\mu_{\rm eq} + \beta_{\alpha}\dd T^{\mu \alpha}_{\rm eq} -  \frac{1}{2} \omega_{\alpha \beta} \dd S^{\mu,\alpha \beta}_{\rm eq},
\eel
\bel{eq:dcalN}
\dd \mathcal{N}^{\mu} = N^{\mu}_{\rm eq}\dd \xi - T^{\mu \alpha}_{\rm eq}\dd \beta_{\alpha} + \frac{1}{2} S^{\mu,\alpha \beta}_{\rm eq}\dd \omega_{\alpha \beta},
\eel
where $S^\mu_{\rm eq} $ is the entropy current. For the classical statistics, one finds that  $\mathcal{N}^{\mu} = (\cosh\xi/\sinh\xi) N^{\mu}$. A direct consequence of \EQ{eq:dSmu} is that the conservation laws~\EQn{eq:psh} imply the entropy conservation $\p_\mu S^\mu_{\rm eq}=0$. One can also notice that~\EQ{eq:dcalN} is not independent but follows from~\EQn{eq:Smu_FD1} and~\EQn{eq:dSmu}.

A popular strategy adopted in the past was to ignore contributions related to $S^{\mu,\alpha \beta}_{\rm eq}$ in~\EQSM{eq:Smu_FD1}{eq:dcalN}~\CITn{Florkowski:2018fap}. This is justified if one considers only the zeroth- and first-order terms in $\omega_{\mu\nu}$. In practice, all calculations suggest that the expansion of $S^{\mu,\alpha \beta}_{\rm eq}$ begins with terms linear in the coefficients of $\omega_{\mu \nu}$, hence non-trivial spin contributions in thermodynamic relations start with quadratic terms. If solely the zeroth- and first-order terms in $\omega_{\mu\nu}$ are included, the thermodynamic relations \EQSMn{eq:Smu_FD1}{eq:dcalN} are reduced to standard scalar forms without spin degrees of freedom. Moreover, in this case, the spin evolution can be treated as taking place in an externally given standard hydrodynamic background. This has allowed studies of spin evolution in simplified~\mbox{\CITn{Wang:2021ngp, Xie:2023gbo,Drogosz:2024lkx}} and very recently also in realistic~\CITn{Singh:2024cub} models of expansion.

It is important to stress at this point that the only expansion parameter discussed so far has been the magnitude of the spin polarization tensor components $\omega_{\mu\nu}$, which is a dimensionless quantity in natural units. The gradient expansion is introduced below in the context of including dissipation.

%%%%%%%%%%%%%%%%%%%%%%%%% main text
\section{Applicability range}
\label{sec:appran}

In both the classical and the Wigner approaches, the key role is played by the scalar density, which can be considered the generating function for all the tensors of interest. In the classical case, 
\begin{equation}\label{ncl}
n_{\rm cl} =\int \dd P \exp\left(- p \cdot \beta \right)
\int \dd S \exp\left(\frac{1}{2} \omega_{\alpha \beta} s^{\alpha \beta} \right),
\end{equation}
and in the quantum case,
\begin{equation}\label{nqt}
n_{\rm qt} = 2 \! \int \! \dd P  \exp\left(- p \cdot \beta \right) \,\cosh\!\sqrt{-\f{1}{4m^2}{\tilde \omega}_{\alpha\beta}(x)p^\beta {\tilde \omega}^\alpha{}_{\mu}(x)p^\mu},
\end{equation}
with the integration measures given by \EQ{eq:dP_dS}.
The spin polarization tensor $\omega_{\mu\nu}$ can be written in terms of its electric- and magneticlike components, $\ev =(e^1, e^2, e^3)$ and \mbox{$\bv=(b^1,b^2,b^3)$}, as
\begin{equation}
\omega_{\mu\nu} = 
\begin{bmatrix}
0     &  e^1 & e^2 & e^3 \\
-e^1  &  0    & -b^3 & b^2 \\
-e^2  &  b^3 & 0 & -b^1 \\
-e^3  & -b^2 & b^1 & 0
\end{bmatrix},
\end{equation}
whereas the dual tensor $\tilde{\omega}_{\mu\nu}$ can be obtained from it via the substitution $\ev \rightarrow \bv$ and $\bv \rightarrow -\ev$.

The convergence of integrals (\ref{ncl}) and (\ref{nqt}) was investigated in the recent article~\CITn{Drogosz:2025ihp}. As the expressions are scalars, they can be written in the local rest frame of the fluid element (LRF), where the transformed quantities are denoted by primes, and $u'^\mu = (1,0,0,0)$. In particular, the three-vectors $\ev$ and $\bv$ transform as
\begin{align}
\label{eq:fieldsJacksone}
\ev^\prime &= \gamma \left( \ev + \vv \times \bv \right) - \frac{\gamma^2}{\gamma+1} \vv (\vv \cdot \ev),\\
\bv^\prime &= \gamma \left( \bv - \vv \times \ev \right) - \frac{\gamma^2}{\gamma+1} \vv (\vv \cdot \bv).\label{eq:fieldsJacksonb}
\end{align}

Furthermore, the integral over the spin space can be performed analytically \cite{Florkowski:2018fap}, yielding in the LRF
\begin{equation}\label{ncl2}
n_{\rm cl} = 2 \int \dd P^\prime   \exp\left(- \f{E_{p^\prime}}{T} \right)
\f{\sinh( \spin |\bv^\prime_*|)}{ \spin \, |\bv^\prime_*|},
\end{equation}
where
\begin{equation}
\bv^\prime_* = \f{1}{m} \left(  E_{p^\prime} \, \bv^\prime - \pv^\prime \times \ev^\prime - \f{\pv^\prime \cdot \bv^\prime}{E_{p^\prime} + m} \pv^\prime \right).
\end{equation}

What determines the convergence of integrals (\ref{nqt}) and~(\ref{ncl2}) is their asymptotic behavior for large magnitudes of $\pvp$. In this regime, both the hyperbolic sine and the hyperbolic cosine behave like the exponential function, and it turns out that the exponents in (\ref{nqt}) and (\ref{ncl2}) differ only by a multiplicative constant and are equal to
\begin{equation}\label{exp}
\hspace{-0.2cm}|\pvp|\!\!\left[-\f {1}{T}
\!+ \! \f{C}{m} \sqrt{{\bvp}^2 \!- \! ({\hat \pv}^\prime \!\cdot\! {\bv}^\prime)^2 \! +\! ({\hat \pv}^\prime \!\times \!\evp)^2 \!-\! 2 {\hat \pv}^\prime \!\cdot \!(\evp \! \times\! \bvp) }\,\right]\!\!,\!\!
\end{equation}
where ${\hat\pv}^\prime$ is the unit vector in the direction of $\pvp$, and $C$ equals $\spin$ in the classical and $\onehalf$ in the quantum case.

The integrals converge if the expression (\ref{exp}) is negative for every direction of the three-vector $\pv$. Maximization of the expression under the square root with respect to ${\hat\pv}^\prime$ results in the applicability criterion 
\begin{equation}\label{criterion}
C \sqrt{ {\bvp}^2 + {\evp}^2 + 2 |\evp \times \bvp|} < \f{m}{T}.
\end{equation}
If one is given only a bound on the absolute value of the components of $\omega'_{\mu\nu}$, then a suitable criterion can be found by noting that for $e'^i, b'^i \in [-\omega'_{\rm max}, \omega'_{\rm max}]$ the expression under the square root in (\ref{exp}) is bounded from above by $\omega'_{\rm max}(2 + \sqrt 2)$. This leads to the criterion
\begin{equation}\label{maxmag}
C \omega_{\rm max}^\prime \left(2 + \sqrt 2\right) < 
\f{m}{T}.
\end{equation}
Further, one can show that for any directions of $\ev$, $\bv$, and $\vv$, $\omega'_{\mu\nu}$ is bounded from above by $\omega'_{\mu\nu}\sqrt{\f{1+v}{1-v}}$. Thus,
\begin{equation}\label{maxmag2}
C \omega_{\rm max} \left(2 + \sqrt{2} \right)\sqrt{\f{1+v}{1-v}} < \f{m}{T}.
\end{equation}
The proofs of the inequalities used to derive (\ref{maxmag}) and (\ref{maxmag2}) are given in \CIT{Drogosz:2025ihp} (see also Ref.~\cite{Abboud:2025qtg} for a recent study).

The criterion (\ref{criterion}) is finer than the next two, which are its upper bounds corresponding to the worst-case choice of values and orientations of the vectors $\ev$, $\bv$, and $\vv$. However, criterion (\ref{maxmag2}) may be the most convenient in practical calculations, as it uses only two variables, $\omega_{\rm max}$ and $v$, rather than the complete set of components of $\ev$, $\bv$, and $\vv$.

%%%%%%%%%%%%%%%%%%%%%%%%% main text
\section{Close-to-equilibrium dynamics}
\label{sec:closeeq}

To include dissipation, we closely follow the method of Israel and Stewart~\CITn{Israel:1979wp} and replace the equilibrium currents $N^\mu_{\rm eq}$, $T^{\mu \alpha}_{\rm eq}$ and $S^{\mu, \alpha \beta}_{\rm eq}$ in \EQ{eq:Smu_FD1} by the general nonequilibrium tensors. The latter can be written as the equilibrium terms plus nonequilibrium corrections: \mbox{$N^\mu = N^\mu_{\rm eq} + \delta N^\mu$}, \mbox{$T^{\mu \alpha} = T^{\mu \alpha}_{\rm eq} + \delta T^{\mu \alpha}$}, \mbox{$S^{\mu, \alpha \beta} = S^{\mu, \alpha \beta}_{\rm eq} +\delta S^{\mu, \alpha \beta}$}. In this way we obtain
\bel{eq:HmuN}
S^\mu =  T^{\mu \alpha} \beta_\alpha-\f{1}{2} \omega_{\alpha\beta} S^{\mu, \alpha \beta}
-\xi N^\mu + {\cal N}^\mu.
\eel 
In the next step, we calculate the entropy production $\p_\mu S^\mu$ using again the conservation laws. The main difference from the description of the perfect fluid is that the spin tensor is no longer conserved. We recall that in the general case, i.e., with dissipation included,  only the total angular momentum is conserved. In this case, the conservation law \mbox{$\p_\mu J^{\mu, \alpha\beta}=0$} with \mbox{$J^{\mu, \alpha\beta} = x^\alpha T^{\mu \beta} - x^\beta T^{\mu \alpha} + S^{\mu, \alpha\beta}$} gives \mbox{$\p_\mu S^{\mu, \alpha\beta}= T^{\beta \alpha}-T^{\alpha \beta}$}. Calculating the divergence of the entropy current defined by \EQ{eq:HmuN} gives
\begin{align}\begin{split}\label{eq:divS}
\p_\mu S^\mu &=
- \delta N^\mu \p_\mu \xi
+ \delta T^{\mu \lambda}_s \p_\mu \beta_\lambda \\
&+ \delta T^{\mu \lambda}_a \LB \p_\mu \beta_\lambda 
- \omega_{\lambda\mu} \RB
-\f{1}{2} \delta S^{\mu, \alpha \beta} \p_\mu \omega_{\alpha\beta} ,
\end{split}\end{align}
where $ \delta T^{\mu \lambda}_s$ and $ \delta T^{\mu \lambda}_a$ are symmetric and antisymmetric dissipative corrections to the energy-momentum tensor, respectively. These and other dissipative terms can be determined from the condition $\p_\mu S^\mu \geq 0$, following the ideas originally introduced in~\CITn{Hattori:2019lfp}. This leads to expressions where $\delta N^\mu$, $\delta T^{\mu \lambda} =  \delta T^{\mu \lambda}_s +  \delta T^{\mu \lambda}_a$, and $\delta S^{\mu, \alpha \beta}$ are expressed as combinations of the gradients of the hydrodynamic variables used in the description of perfect fluid. 

The strategy outlined above clearly suggests that our approach is based on a two-fold expansion, both in the magnitude of $\omega_{\mu\nu}$ and in the gradients of the hydrodynamic variables. As long as we remain at the perfect-fluid level, we consider an expansion in $\omega_{\alpha\beta}$ only. If dissipation is included, we have to include in parallel the gradient expansion. This approach is similar to the treatment of the baryon chemical potential $\mu$ in standard hydrodynamics -- one can always consider an expansion in the magnitude of $\xi =\mu/T$, together with the gradient expansion. 

Although \EQ{eq:divS} or its special case for $\xi=0$ was obtained before (see, for example: Eq.~(10) in \CITn{Hattori:2019lfp}, (23) in \CITn{Fukushima:2020ucl}, (21) in \CITn{Biswas:2023qsw}, and the QFT analysis in \CITn{Becattini:2023ouz}), the previous studies considered always the local equilibrium state without the second-order corrections discussed above. Thus, it is important to extend the previous analyses by considering a different reference point for local equilibrium quantities. 

Equation \EQn{eq:divS} implies that the {\it global equilibrium} is defined by the generalized Tolman-Klein conditions~\CITn{Tolman:1934,Klein:1949}, which include the two standard equations, $\p_\mu \xi = 0$ and $\p_{( \mu} \beta_{\lambda )} = 0$, and an extra constraint that the spin polarization tensor is given by the thermal vorticity, $\omega_{\lambda\mu} = \p_{[ \mu} \beta_{\lambda ]}$, where the indices between round (square) brackets are to be symmetrized (antisymmetrized) over; conf.~Eqs.~(\ref{thvort})-(\ref{thsh}). Nevertheless, in {\it local equilibrium} $\omega_{\lambda\mu}$ and $\p_{[ \mu} \beta_{\lambda ]}$ are not directly related and may be significantly different from each other. In this respect, we differ from the concept of local equilibrium originally proposed in~\CITn{Becattini:2013fla}. As stated above, the behavior of the spin polarization tensor, $\omega_{\mu\nu} = \Omega_{\mu\nu}/T$, is similar to the behavior of the ratio $\xi = \mu/T$ in standard (spinless) relativistic hydrodynamics. In global equilibrium $\xi =$~const., while in local equilibrium a direct connection between $T$ and $\mu$ is lost. 

%%%%%%%%%%%%%%%%%%%%%%%%% main text
\section{Summary and outlook}
\label{sec:summary}

Following Refs.~\CITn{Florkowski:2024bfw, Drogosz:2024gzv}, we propose to use a hybrid form of spin hydrodynamics that combines the most attractive features of various descriptions of spin hydrodynamics that have been developed so far. The hybrid approach allows us to establish relations and connections between the known frameworks and provides solutions to several important problems:

\medskip \noindent
{\bf  (i)} Expansion up to the second order in the spin polarization tensor $\omega_{\alpha\beta}$ offers us an opportunity to maintain nontrivial spin thermodynamic relations without the contradictory assumption that the spin tensor and the spin polarization tensor are of different orders. 

\medskip \noindent
{\bf (ii)} The stability and causality problems encountered in the frameworks using the Israel-Stewart method can be solved by referring to the kinetic-theory result that explains the different dependence of the spin tensor on the electric and magnetic components~\CITn{Daher:2024ixz}. 

\medskip \noindent
{\bf  (iii)} The two-fold character of the expansion described above allows us to disentangle different phenomena that otherwise seem to be connected. For example, since in dissipative spin hydrodynamics one finds the combination $\beta a^\mu + \beta^2 \nabla^\mu T - 2 \omega^{\mu\nu} u_\nu$ (where $\beta$ is the inverse temperature, $a^\mu$ the four acceleration, and $\nabla^\mu$ the transverse gradient), in the literature it is popular to assume that $\omega_{\mu\nu}$ has the character of a gradient term. There is no reason to do so in our approach.  

\medskip \noindent
{\bf  (iv)} In our recent work~\CITn{Bhadury:2025boe} we have proposed an improved version of the local equilibrium Wigner function and argued that it should replace the formula given in~\CITn{Becattini:2013fla}. We have demonstrated that the framework of~\CITn{Bhadury:2025boe} agrees with that based on the classical description of spin up to quadratic terms in the spin polarization tensor. 

\medskip \noindent
{\bf  (v)} We have determined the applicability range of perfect spin hydrodynamics~\CITn{Drogosz:2025ihp} and found it compatible with the conditions existing in the late stages of heavy-ion collisions. 

\medskip
Finally, we emphasize that our approach defines local equilibrium for particles with spin in simple physical terms as the phase dominated by the $s$-wave scattering (alternatively, as the phase where spin changing processes are slow). The inclusion of dissipation relaxes this assumption and allows for processes that change the orbital and spin parts of the total angular momentum. Certainly, it remains to be seen whether such a scheme turns out to be useful and reliable in physical applications to interpret the data. The first, encouraging steps in this direction have been taken in~\CIT{Singh:2024cub}.

%%%%%%%%%%%%%%%%%%%%%%%%%%%%%%%%%%%%%%%%%%%%%%
\bigskip
\noindent
{\it Acknowledgments.} \newline This work was supported in part by the Polish National Science Centre (NCN) Grant No. 2022/47/B/ST2/01372. S.~B. would like to acknowledge the support of the Faculty of Physics, Astronomy, and Applied Computer Science, Jagiellonian University via
Grant No. LM/36/BS.

\bibliographystyle{elsarticle-num}
\bibliography{shrd}

\begin{thebibliography}{10}
\expandafter\ifx\csname url\endcsname\relax
  \def\url#1{\texttt{#1}}\fi
\expandafter\ifx\csname urlprefix\endcsname\relax\def\urlprefix{URL }\fi
\expandafter\ifx\csname href\endcsname\relax
  \def\href#1#2{#2} \def\path#1{#1}\fi

\bibitem{Busza:2018rrf}
W.~Busza, K.~Rajagopal, W.~van~der Schee, {Heavy Ion Collisions: The Big Picture, and the Big Questions}, Ann. Rev. Nucl. Part. Sci. 68 (2018) 339--376.
\newblock \href {http://arxiv.org/abs/1802.04801} {\path{arXiv:1802.04801}}, \href {https://doi.org/10.1146/annurev-nucl-101917-020852} {\path{doi:10.1146/annurev-nucl-101917-020852}}.

\bibitem{Romatschke:2007mq}
P.~Romatschke, U.~Romatschke, {Viscosity Information from Relativistic Nuclear Collisions: How Perfect is the Fluid Observed at RHIC?}, Phys. Rev. Lett. 99 (2007) 172301.
\newblock \href {http://arxiv.org/abs/0706.1522} {\path{arXiv:0706.1522}}, \href {https://doi.org/10.1103/PhysRevLett.99.172301} {\path{doi:10.1103/PhysRevLett.99.172301}}.

\bibitem{Romatschke:2009im}
P.~Romatschke, {New Developments in Relativistic Viscous Hydrodynamics}, Int. J. Mod. Phys. E 19 (2010) 1--53.
\newblock \href {http://arxiv.org/abs/0902.3663} {\path{arXiv:0902.3663}}, \href {https://doi.org/10.1142/S0218301310014613} {\path{doi:10.1142/S0218301310014613}}.

\bibitem{Jaiswal:2016hex}
A.~Jaiswal, V.~Roy, {Relativistic hydrodynamics in heavy-ion collisions: general aspects and recent developments}, Adv. High Energy Phys. 2016 (2016) 9623034.
\newblock \href {http://arxiv.org/abs/1605.08694} {\path{arXiv:1605.08694}}, \href {https://doi.org/10.1155/2016/9623034} {\path{doi:10.1155/2016/9623034}}.

\bibitem{Florkowski:2017olj}
W.~Florkowski, M.~P. Heller, M.~Spalinski, {New theories of relativistic hydrodynamics in the LHC era}, Rept. Prog. Phys. 81~(4) (2018) 046001.
\newblock \href {http://arxiv.org/abs/1707.02282} {\path{arXiv:1707.02282}}, \href {https://doi.org/10.1088/1361-6633/aaa091} {\path{doi:10.1088/1361-6633/aaa091}}.

\bibitem{landau1987fluid}
L.~Landau, E.~Lifshitz, Fluid mechanics., Vol.~6, Elsevier, 1987.

\bibitem{Eckart:1940te}
C.~Eckart, {The Thermodynamics of irreversible processes. 3.. Relativistic theory of the simple fluid}, Phys. Rev. 58 (1940) 919--924.
\newblock \href {https://doi.org/10.1103/PhysRev.58.919} {\path{doi:10.1103/PhysRev.58.919}}.

\bibitem{Hiscock:1983zz}
W.~A. Hiscock, L.~Lindblom, {Stability and causality in dissipative relativistic fluids}, Annals Phys. 151 (1983) 466--496.
\newblock \href {https://doi.org/10.1016/0003-4916(83)90288-9} {\path{doi:10.1016/0003-4916(83)90288-9}}.

\bibitem{Hiscock:1985zz}
W.~A. Hiscock, L.~Lindblom, {Generic instabilities in first-order dissipative relativistic fluid theories}, Phys. Rev. D 31 (1985) 725--733.
\newblock \href {https://doi.org/10.1103/PhysRevD.31.725} {\path{doi:10.1103/PhysRevD.31.725}}.

\bibitem{Muller:1967zza}
I.~Muller, {Zum Paradoxon der Warmeleitungstheorie}, Z. Phys. 198 (1967) 329--344.
\newblock \href {https://doi.org/10.1007/BF01326412} {\path{doi:10.1007/BF01326412}}.

\bibitem{Israel:1976tn}
W.~Israel, {Nonstationary irreversible thermodynamics: A Causal relativistic theory}, Annals Phys. 100 (1976) 310--331.
\newblock \href {https://doi.org/10.1016/0003-4916(76)90064-6} {\path{doi:10.1016/0003-4916(76)90064-6}}.

\bibitem{Israel:1979wp}
W.~Israel, J.~M. Stewart, {Transient relativistic thermodynamics and kinetic theory}, Annals Phys. 118 (1979) 341--372.
\newblock \href {https://doi.org/10.1016/0003-4916(79)90130-1} {\path{doi:10.1016/0003-4916(79)90130-1}}.

\bibitem{Baier:2007ix}
R.~Baier, P.~Romatschke, D.~T. Son, A.~O. Starinets, M.~A. Stephanov, {Relativistic viscous hydrodynamics, conformal invariance, and holography}, JHEP 04 (2008) 100.
\newblock \href {http://arxiv.org/abs/0712.2451} {\path{arXiv:0712.2451}}, \href {https://doi.org/10.1088/1126-6708/2008/04/100} {\path{doi:10.1088/1126-6708/2008/04/100}}.

\bibitem{Denicol:2012cn}
G.~S. Denicol, H.~Niemi, E.~Molnar, D.~H. Rischke, {Derivation of transient relativistic fluid dynamics from the Boltzmann equation}, Phys. Rev. D 85 (2012) 114047, [Erratum: Phys.Rev.D 91, 039902 (2015)].
\newblock \href {http://arxiv.org/abs/1202.4551} {\path{arXiv:1202.4551}}, \href {https://doi.org/10.1103/PhysRevD.85.114047} {\path{doi:10.1103/PhysRevD.85.114047}}.

\bibitem{Florkowski:2010cf}
W.~Florkowski, R.~Ryblewski, {Highly-anisotropic and strongly-dissipative hydrodynamics for early stages of relativistic heavy-ion collisions}, Phys. Rev. C 83 (2011) 034907.
\newblock \href {http://arxiv.org/abs/1007.0130} {\path{arXiv:1007.0130}}, \href {https://doi.org/10.1103/PhysRevC.83.034907} {\path{doi:10.1103/PhysRevC.83.034907}}.

\bibitem{Martinez:2010sc}
M.~Martinez, M.~Strickland, {Dissipative Dynamics of Highly Anisotropic Systems}, Nucl. Phys. A 848 (2010) 183--197.
\newblock \href {http://arxiv.org/abs/1007.0889} {\path{arXiv:1007.0889}}, \href {https://doi.org/10.1016/j.nuclphysa.2010.08.011} {\path{doi:10.1016/j.nuclphysa.2010.08.011}}.

\bibitem{El:2009vj}
A.~El, Z.~Xu, C.~Greiner, {Third-order relativistic dissipative hydrodynamics}, Phys. Rev. C 81 (2010) 041901.
\newblock \href {http://arxiv.org/abs/0907.4500} {\path{arXiv:0907.4500}}, \href {https://doi.org/10.1103/PhysRevC.81.041901} {\path{doi:10.1103/PhysRevC.81.041901}}.

\bibitem{Jaiswal:2013vta}
A.~Jaiswal, {Relativistic third-order dissipative fluid dynamics from kinetic theory}, Phys. Rev. C 88 (2013) 021903.
\newblock \href {http://arxiv.org/abs/1305.3480} {\path{arXiv:1305.3480}}, \href {https://doi.org/10.1103/PhysRevC.88.021903} {\path{doi:10.1103/PhysRevC.88.021903}}.

\bibitem{Bemfica:2017wps}
F.~S. Bemfica, M.~M. Disconzi, J.~Noronha, {Causality and existence of solutions of relativistic viscous fluid dynamics with gravity}, Phys. Rev. D 98~(10) (2018) 104064.
\newblock \href {http://arxiv.org/abs/1708.06255} {\path{arXiv:1708.06255}}, \href {https://doi.org/10.1103/PhysRevD.98.104064} {\path{doi:10.1103/PhysRevD.98.104064}}.

\bibitem{Kovtun:2019hdm}
P.~Kovtun, {First-order relativistic hydrodynamics is stable}, JHEP 10 (2019) 034.
\newblock \href {http://arxiv.org/abs/1907.08191} {\path{arXiv:1907.08191}}, \href {https://doi.org/10.1007/JHEP10(2019)034} {\path{doi:10.1007/JHEP10(2019)034}}.

\bibitem{Bemfica:2019knx}
F.~S. Bemfica, F.~S. Bemfica, M.~M. Disconzi, M.~M. Disconzi, J.~Noronha, J.~Noronha, {Nonlinear Causality of General First-Order Relativistic Viscous Hydrodynamics}, Phys. Rev. D 100~(10) (2019) 104020, [Erratum: Phys.Rev.D 105, 069902 (2022)].
\newblock \href {http://arxiv.org/abs/1907.12695} {\path{arXiv:1907.12695}}, \href {https://doi.org/10.1103/PhysRevD.100.104020} {\path{doi:10.1103/PhysRevD.100.104020}}.

\bibitem{Armas:2020mpr}
J.~Armas, A.~Jain, {Effective field theory for hydrodynamics without boosts}, SciPost Phys. 11~(3) (2021) 054.
\newblock \href {http://arxiv.org/abs/2010.15782} {\path{arXiv:2010.15782}}, \href {https://doi.org/10.21468/SciPostPhys.11.3.054} {\path{doi:10.21468/SciPostPhys.11.3.054}}.

\bibitem{Basar:2024qxd}
G.~Ba\c{s}ar, J.~Bhambure, R.~Singh, D.~Teaney, {Stochastic relativistic advection diffusion equation~from the Metropolis algorithm}, Phys. Rev. C 110~(4) (2024) 044903.
\newblock \href {http://arxiv.org/abs/2403.04185} {\path{arXiv:2403.04185}}, \href {https://doi.org/10.1103/PhysRevC.110.044903} {\path{doi:10.1103/PhysRevC.110.044903}}.

\bibitem{Bhattacharyya:2024ohn}
S.~Bhattacharyya, S.~Mitra, S.~Roy, R.~Singh, {Field redefinition and its impact in relativistic hydrodynamics}, Phys. Rev. D 111~(1) (2025) 014034.
\newblock \href {http://arxiv.org/abs/2409.15387} {\path{arXiv:2409.15387}}, \href {https://doi.org/10.1103/PhysRevD.111.014034} {\path{doi:10.1103/PhysRevD.111.014034}}.

\bibitem{Denicol:2018rbw}
G.~S. Denicol, X.-G. Huang, E.~Moln\'ar, G.~M. Monteiro, H.~Niemi, J.~Noronha, D.~H. Rischke, Q.~Wang, {Nonresistive dissipative magnetohydrodynamics from the Boltzmann equation in the 14-moment approximation}, Phys. Rev. D 98~(7) (2018) 076009.
\newblock \href {http://arxiv.org/abs/1804.05210} {\path{arXiv:1804.05210}}, \href {https://doi.org/10.1103/PhysRevD.98.076009} {\path{doi:10.1103/PhysRevD.98.076009}}.

\bibitem{Panda:2020zhr}
A.~K. Panda, A.~Dash, R.~Biswas, V.~Roy, {Relativistic non-resistive viscous magnetohydrodynamics from the kinetic theory: a relaxation time approach}, JHEP 03 (2021) 216.
\newblock \href {http://arxiv.org/abs/2011.01606} {\path{arXiv:2011.01606}}, \href {https://doi.org/10.1007/JHEP03(2021)216} {\path{doi:10.1007/JHEP03(2021)216}}.

\bibitem{Stephanov:2017ghc}
M.~Stephanov, Y.~Yin, {Hydrodynamics with parametric slowing down and fluctuations near the critical point}, Phys. Rev. D 98~(3) (2018) 036006.
\newblock \href {http://arxiv.org/abs/1712.10305} {\path{arXiv:1712.10305}}, \href {https://doi.org/10.1103/PhysRevD.98.036006} {\path{doi:10.1103/PhysRevD.98.036006}}.

\bibitem{An:2019osr}
X.~An, G.~Basar, M.~Stephanov, H.-U. Yee, {Relativistic Hydrodynamic Fluctuations}, Phys. Rev. C 100~(2) (2019) 024910.
\newblock \href {http://arxiv.org/abs/1902.09517} {\path{arXiv:1902.09517}}, \href {https://doi.org/10.1103/PhysRevC.100.024910} {\path{doi:10.1103/PhysRevC.100.024910}}.

\bibitem{STAR:2017ckg}
L.~Adamczyk, et~al., {Global $\Lambda$ hyperon polarization in nuclear collisions: evidence for the most vortical fluid}, Nature 548 (2017) 62--65.
\newblock \href {http://arxiv.org/abs/1701.06657} {\path{arXiv:1701.06657}}, \href {https://doi.org/10.1038/nature23004} {\path{doi:10.1038/nature23004}}.

\bibitem{Adam:2018ivw}
J.~Adam, L.~Adamczyk, J.~Adams, J.~K. Adkins, G.~Agakishiev, M.~Aggarwal, Z.~Ahammed, N.~Ajitanand, I.~Alekseev, D.~Anderson, et~al., Global polarization of $\lambda$ hyperons in au+ au collisions at s nn= 200 gev, Physical Review C 98~(1) (2018) 014910.

\bibitem{Niida:2018hfw}
T.~Niida, {Global and local polarization of $\Lambda$ hyperons in Au+Au collisions at 200 GeV from STAR}, Nucl. Phys. A 982 (2019) 511--514.
\newblock \href {http://arxiv.org/abs/1808.10482} {\path{arXiv:1808.10482}}, \href {https://doi.org/10.1016/j.nuclphysa.2018.08.034} {\path{doi:10.1016/j.nuclphysa.2018.08.034}}.

\bibitem{Romatschke:2017ejr}
P.~Romatschke, U.~Romatschke, Relativistic fluid dynamics in and out of equilibrium—ten years of progress in theory and numerical simulations of nuclear collisions, arXiv preprint arXiv:1712.05815 (2017).

\bibitem{Becattini:2013fla}
F.~Becattini, V.~Chandra, L.~Del~Zanna, E.~Grossi, {Relativistic distribution function for particles with spin at local thermodynamical equilibrium}, Annals Phys. 338 (2013) 32--49.
\newblock \href {http://arxiv.org/abs/1303.3431} {\path{arXiv:1303.3431}}, \href {https://doi.org/10.1016/j.aop.2013.07.004} {\path{doi:10.1016/j.aop.2013.07.004}}.

\bibitem{Becattini:2021iol}
F.~Becattini, M.~Buzzegoli, G.~Inghirami, I.~Karpenko, A.~Palermo, {Local Polarization and Isothermal Local Equilibrium in Relativistic Heavy Ion Collisions}, Phys. Rev. Lett. 127~(27) (2021) 272302.
\newblock \href {http://arxiv.org/abs/2103.14621} {\path{arXiv:2103.14621}}, \href {https://doi.org/10.1103/PhysRevLett.127.272302} {\path{doi:10.1103/PhysRevLett.127.272302}}.

\bibitem{Palermo:2024tza}
A.~Palermo, E.~Grossi, I.~Karpenko, F.~Becattini, {$\Lambda $ polarization in very high energy heavy ion collisions as a probe of the quark\textendash{}gluon plasma formation and properties}, Eur. Phys. J. C 84~(9) (2024) 920.
\newblock \href {http://arxiv.org/abs/2404.14295} {\path{arXiv:2404.14295}}, \href {https://doi.org/10.1140/epjc/s10052-024-13229-z} {\path{doi:10.1140/epjc/s10052-024-13229-z}}.

\bibitem{Florkowski:2017ruc}
W.~Florkowski, B.~Friman, A.~Jaiswal, E.~Speranza, {Relativistic fluid dynamics with spin}, Phys. Rev. C97~(4) (2018) 041901.
\newblock \href {http://arxiv.org/abs/1705.00587} {\path{arXiv:1705.00587}}, \href {https://doi.org/10.1103/PhysRevC.97.041901} {\path{doi:10.1103/PhysRevC.97.041901}}.

\bibitem{Florkowski:2017dyn}
W.~Florkowski, B.~Friman, A.~Jaiswal, R.~Ryblewski, E.~Speranza, {Spin-dependent distribution functions for relativistic hydrodynamics of spin-1/2 particles}, Phys. Rev. D 97~(11) (2018) 116017.
\newblock \href {http://arxiv.org/abs/1712.07676} {\path{arXiv:1712.07676}}, \href {https://doi.org/10.1103/PhysRevD.97.116017} {\path{doi:10.1103/PhysRevD.97.116017}}.

\bibitem{Florkowski:2018ahw}
W.~Florkowski, A.~Kumar, R.~Ryblewski, {Thermodynamic versus kinetic approach to polarization-vorticity coupling}, Phys. Rev. C 98~(4) (2018) 044906.
\newblock \href {http://arxiv.org/abs/1806.02616} {\path{arXiv:1806.02616}}, \href {https://doi.org/10.1103/PhysRevC.98.044906} {\path{doi:10.1103/PhysRevC.98.044906}}.

\bibitem{Florkowski:2018fap}
W.~Florkowski, A.~Kumar, R.~Ryblewski, {Relativistic hydrodynamics for spin-polarized fluids}, Prog. Part. Nucl. Phys. 108 (2019) 103709.
\newblock \href {http://arxiv.org/abs/1811.04409} {\path{arXiv:1811.04409}}, \href {https://doi.org/10.1016/j.ppnp.2019.07.001} {\path{doi:10.1016/j.ppnp.2019.07.001}}.

\bibitem{Bhadury:2020puc}
S.~Bhadury, W.~Florkowski, A.~Jaiswal, A.~Kumar, R.~Ryblewski, {Relativistic dissipative spin dynamics in the relaxation time approximation}, Phys. Lett. B 814 (2021) 136096.
\newblock \href {http://arxiv.org/abs/2002.03937} {\path{arXiv:2002.03937}}, \href {https://doi.org/10.1016/j.physletb.2021.136096} {\path{doi:10.1016/j.physletb.2021.136096}}.

\bibitem{Bhadury:2022ulr}
S.~Bhadury, W.~Florkowski, A.~Jaiswal, A.~Kumar, R.~Ryblewski, {Relativistic Spin Magnetohydrodynamics}, Phys. Rev. Lett. 129~(19) (2022) 192301.
\newblock \href {http://arxiv.org/abs/2204.01357} {\path{arXiv:2204.01357}}, \href {https://doi.org/10.1103/PhysRevLett.129.192301} {\path{doi:10.1103/PhysRevLett.129.192301}}.

\bibitem{Weickgenannt:2019dks}
N.~Weickgenannt, X.-L. Sheng, E.~Speranza, Q.~Wang, D.~H. Rischke, {Kinetic theory for massive spin-1/2 particles from the Wigner-function formalism}, Phys. Rev. D 100~(5) (2019) 056018.
\newblock \href {http://arxiv.org/abs/1902.06513} {\path{arXiv:1902.06513}}, \href {https://doi.org/10.1103/PhysRevD.100.056018} {\path{doi:10.1103/PhysRevD.100.056018}}.

\bibitem{Weickgenannt:2021cuo}
N.~Weickgenannt, E.~Speranza, X.-l. Sheng, Q.~Wang, D.~H. Rischke, {Derivation of the nonlocal collision term in the relativistic Boltzmann equation for massive spin-1/2 particles from quantum field theory}, Phys. Rev. D 104~(1) (2021) 016022.
\newblock \href {http://arxiv.org/abs/2103.04896} {\path{arXiv:2103.04896}}, \href {https://doi.org/10.1103/PhysRevD.104.016022} {\path{doi:10.1103/PhysRevD.104.016022}}.

\bibitem{Weickgenannt:2020aaf}
N.~Weickgenannt, E.~Speranza, X.-l. Sheng, Q.~Wang, D.~H. Rischke, {Generating Spin Polarization from Vorticity through Nonlocal Collisions}, Phys. Rev. Lett. 127~(5) (2021) 052301.
\newblock \href {http://arxiv.org/abs/2005.01506} {\path{arXiv:2005.01506}}, \href {https://doi.org/10.1103/PhysRevLett.127.052301} {\path{doi:10.1103/PhysRevLett.127.052301}}.

\bibitem{Weickgenannt:2022zxs}
N.~Weickgenannt, D.~Wagner, E.~Speranza, D.~H. Rischke, {Relativistic second-order dissipative spin hydrodynamics from the method of moments}, Phys. Rev. D 106~(9) (2022) 096014.
\newblock \href {http://arxiv.org/abs/2203.04766} {\path{arXiv:2203.04766}}, \href {https://doi.org/10.1103/PhysRevD.106.096014} {\path{doi:10.1103/PhysRevD.106.096014}}.

\bibitem{Weickgenannt:2023nge}
N.~Weickgenannt, J.-P. Blaizot, {Chiral hydrodynamics of expanding systems}, Phys. Rev. D 109~(5) (2024) 056012.
\newblock \href {http://arxiv.org/abs/2311.15817} {\path{arXiv:2311.15817}}, \href {https://doi.org/10.1103/PhysRevD.109.056012} {\path{doi:10.1103/PhysRevD.109.056012}}.

\bibitem{Wagner:2024fhf}
D.~Wagner, M.~Shokri, D.~H. Rischke, {Damping of spin waves}, Phys. Rev. Res. 6~(4) (2024) 043103.
\newblock \href {http://arxiv.org/abs/2405.00533} {\path{arXiv:2405.00533}}, \href {https://doi.org/10.1103/PhysRevResearch.6.043103} {\path{doi:10.1103/PhysRevResearch.6.043103}}.

\bibitem{Hu:2021pwh}
J.~Hu, {Relativistic first-order spin hydrodynamics via the Chapman-Enskog expansion}, Phys. Rev. D 105~(7) (2022) 076009.
\newblock \href {http://arxiv.org/abs/2111.03571} {\path{arXiv:2111.03571}}, \href {https://doi.org/10.1103/PhysRevD.105.076009} {\path{doi:10.1103/PhysRevD.105.076009}}.

\bibitem{Li:2020eon}
S.~Li, M.~A. Stephanov, H.-U. Yee, {Nondissipative Second-Order Transport, Spin, and Pseudogauge Transformations in Hydrodynamics}, Phys. Rev. Lett. 127~(8) (2021) 082302.
\newblock \href {http://arxiv.org/abs/2011.12318} {\path{arXiv:2011.12318}}, \href {https://doi.org/10.1103/PhysRevLett.127.082302} {\path{doi:10.1103/PhysRevLett.127.082302}}.

\bibitem{Shi:2020htn}
S.~Shi, C.~Gale, S.~Jeon, {From chiral kinetic theory to relativistic viscous spin hydrodynamics}, Phys. Rev. C 103~(4) (2021) 044906.
\newblock \href {http://arxiv.org/abs/2008.08618} {\path{arXiv:2008.08618}}, \href {https://doi.org/10.1103/PhysRevC.103.044906} {\path{doi:10.1103/PhysRevC.103.044906}}.

\bibitem{Banerjee:2024xnd}
S.~Banerjee, S.~Bhadury, W.~Florkowski, A.~Jaiswal, R.~Ryblewski, {Longitudinal spin polarization in a thermal model with dissipative corrections} (5 2024).
\newblock \href {http://arxiv.org/abs/2405.05089} {\path{arXiv:2405.05089}}.

\bibitem{Bhadury:2024ckc}
S.~Bhadury, {Relativistic spin hydrodynamics with momentum- and spin-dependent relaxation time}, Phys. Rev. C 111~(3) (2025) 034909.
\newblock \href {http://arxiv.org/abs/2408.14462} {\path{arXiv:2408.14462}}, \href {https://doi.org/10.1103/PhysRevC.111.034909} {\path{doi:10.1103/PhysRevC.111.034909}}.

\bibitem{Hattori:2019lfp}
K.~Hattori, M.~Hongo, X.-G. Huang, M.~Matsuo, H.~Taya, {Fate of spin polarization in a relativistic fluid: An entropy-current analysis}, Phys. Lett. B 795 (2019) 100--106.
\newblock \href {http://arxiv.org/abs/1901.06615} {\path{arXiv:1901.06615}}, \href {https://doi.org/10.1016/j.physletb.2019.05.040} {\path{doi:10.1016/j.physletb.2019.05.040}}.

\bibitem{Fukushima:2020ucl}
K.~Fukushima, S.~Pu, {Spin hydrodynamics and symmetric energy-momentum tensors \textendash{} A current induced by the spin vorticity \textendash{}}, Phys. Lett. B 817 (2021) 136346.
\newblock \href {http://arxiv.org/abs/2010.01608} {\path{arXiv:2010.01608}}, \href {https://doi.org/10.1016/j.physletb.2021.136346} {\path{doi:10.1016/j.physletb.2021.136346}}.

\bibitem{Daher:2022xon}
A.~Daher, A.~Das, W.~Florkowski, R.~Ryblewski, {Canonical and phenomenological formulations of spin hydrodynamics}, Phys. Rev. C 108~(2) (2023) 024902.
\newblock \href {http://arxiv.org/abs/2202.12609} {\path{arXiv:2202.12609}}, \href {https://doi.org/10.1103/PhysRevC.108.024902} {\path{doi:10.1103/PhysRevC.108.024902}}.

\bibitem{Daher:2022wzf}
A.~Daher, A.~Das, R.~Ryblewski, {Stability studies of first-order spin-hydrodynamic frameworks}, Phys. Rev. D 107~(5) (2023) 054043.
\newblock \href {http://arxiv.org/abs/2209.10460} {\path{arXiv:2209.10460}}, \href {https://doi.org/10.1103/PhysRevD.107.054043} {\path{doi:10.1103/PhysRevD.107.054043}}.

\bibitem{Sarwar:2022yzs}
G.~Sarwar, M.~Hasanujjaman, J.~R. Bhatt, H.~Mishra, J.-e. Alam, {Causality and stability of relativistic spin hydrodynamics}, Phys. Rev. D 107~(5) (2023) 054031.
\newblock \href {http://arxiv.org/abs/2209.08652} {\path{arXiv:2209.08652}}, \href {https://doi.org/10.1103/PhysRevD.107.054031} {\path{doi:10.1103/PhysRevD.107.054031}}.

\bibitem{Wang:2021ngp}
D.-L. Wang, S.~Fang, S.~Pu, {Analytic solutions of relativistic dissipative spin hydrodynamics with Bjorken expansion}, Phys. Rev. D 104~(11) (2021) 114043.
\newblock \href {http://arxiv.org/abs/2107.11726} {\path{arXiv:2107.11726}}, \href {https://doi.org/10.1103/PhysRevD.104.114043} {\path{doi:10.1103/PhysRevD.104.114043}}.

\bibitem{Biswas:2023qsw}
R.~Biswas, A.~Daher, A.~Das, W.~Florkowski, R.~Ryblewski, {Relativistic second-order spin hydrodynamics: An entropy-current analysis}, Phys. Rev. D 108~(1) (2023) 014024.
\newblock \href {http://arxiv.org/abs/2304.01009} {\path{arXiv:2304.01009}}, \href {https://doi.org/10.1103/PhysRevD.108.014024} {\path{doi:10.1103/PhysRevD.108.014024}}.

\bibitem{Xie:2023gbo}
X.-Q. Xie, D.-L. Wang, C.~Yang, S.~Pu, {Causality and stability analysis for the minimal causal spin hydrodynamics}, Phys. Rev. D 108~(9) (2023) 094031.
\newblock \href {http://arxiv.org/abs/2306.13880} {\path{arXiv:2306.13880}}, \href {https://doi.org/10.1103/PhysRevD.108.094031} {\path{doi:10.1103/PhysRevD.108.094031}}.

\bibitem{Daher:2024ixz}
A.~Daher, W.~Florkowski, R.~Ryblewski, {Stability constraint for spin equation of state}, Phys. Rev. D 110~(3) (2024) 034029.
\newblock \href {http://arxiv.org/abs/2401.07608} {\path{arXiv:2401.07608}}, \href {https://doi.org/10.1103/PhysRevD.110.034029} {\path{doi:10.1103/PhysRevD.110.034029}}.

\bibitem{Ren:2024pur}
X.~Ren, C.~Yang, D.-L. Wang, S.~Pu, {Thermodynamic stability in relativistic viscous and spin hydrodynamics}, Phys. Rev. D 110~(3) (2024) 034010.
\newblock \href {http://arxiv.org/abs/2405.03105} {\path{arXiv:2405.03105}}, \href {https://doi.org/10.1103/PhysRevD.110.034010} {\path{doi:10.1103/PhysRevD.110.034010}}.

\bibitem{Daher:2024bah}
A.~Daher, W.~Florkowski, R.~Ryblewski, F.~Taghinavaz, {Stability and causality of rest frame modes in second-order spin hydrodynamics}, Phys. Rev. D 109~(11) (2024) 114001.
\newblock \href {http://arxiv.org/abs/2403.04711} {\path{arXiv:2403.04711}}, \href {https://doi.org/10.1103/PhysRevD.109.114001} {\path{doi:10.1103/PhysRevD.109.114001}}.

\bibitem{Gallegos:2021bzp}
A.~D. Gallegos, U.~G\"ursoy, A.~Yarom, {Hydrodynamics of spin currents}, SciPost Phys. 11 (2021) 041.
\newblock \href {http://arxiv.org/abs/2101.04759} {\path{arXiv:2101.04759}}, \href {https://doi.org/10.21468/SciPostPhys.11.2.041} {\path{doi:10.21468/SciPostPhys.11.2.041}}.

\bibitem{Hongo:2021ona}
M.~Hongo, X.-G. Huang, M.~Kaminski, M.~Stephanov, H.-U. Yee, {Relativistic spin hydrodynamics with torsion and linear response theory for spin relaxation}, JHEP 11 (2021) 150.
\newblock \href {http://arxiv.org/abs/2107.14231} {\path{arXiv:2107.14231}}, \href {https://doi.org/10.1007/JHEP11(2021)150} {\path{doi:10.1007/JHEP11(2021)150}}.

\bibitem{Kumar:2023ojl}
A.~Kumar, D.-L. Yang, P.~Gubler, {Spin alignment of vector mesons by second-order hydrodynamic gradients}, Phys. Rev. D 109~(5) (2024) 054038.
\newblock \href {http://arxiv.org/abs/2312.16900} {\path{arXiv:2312.16900}}, \href {https://doi.org/10.1103/PhysRevD.109.054038} {\path{doi:10.1103/PhysRevD.109.054038}}.

\bibitem{She:2021lhe}
D.~She, A.~Huang, D.~Hou, J.~Liao, {Relativistic viscous hydrodynamics with angular momentum}, Sci. Bull. 67 (2022) 2265--2268.
\newblock \href {http://arxiv.org/abs/2105.04060} {\path{arXiv:2105.04060}}, \href {https://doi.org/10.1016/j.scib.2022.10.020} {\path{doi:10.1016/j.scib.2022.10.020}}.

\bibitem{Montenegro:2017rbu}
D.~Montenegro, L.~Tinti, G.~Torrieri, {Ideal relativistic fluid limit for a medium with polarization}, Phys. Rev. D 96~(5) (2017) 056012, [Addendum: Phys.Rev.D 96, 079901 (2017)].
\newblock \href {http://arxiv.org/abs/1701.08263} {\path{arXiv:1701.08263}}, \href {https://doi.org/10.1103/PhysRevD.96.056012} {\path{doi:10.1103/PhysRevD.96.056012}}.

\bibitem{Huang:2024ffg}
X.-G. Huang, {An introduction to relativistic spin hydrodynamics} (11 2024).
\newblock \href {http://arxiv.org/abs/2411.11753} {\path{arXiv:2411.11753}}.

\bibitem{Florkowski:2024bfw}
W.~Florkowski, M.~Hontarenko, {Generalized Thermodynamic Relations for Perfect Spin Hydrodynamics}, Phys. Rev. Lett. 134~(8) (2025) 082302.
\newblock \href {http://arxiv.org/abs/2405.03263} {\path{arXiv:2405.03263}}, \href {https://doi.org/10.1103/PhysRevLett.134.082302} {\path{doi:10.1103/PhysRevLett.134.082302}}.

\bibitem{Drogosz:2024gzv}
Z.~Drogosz, W.~Florkowski, M.~Hontarenko, {Hybrid approach to perfect and dissipative spin hydrodynamics}, Phys. Rev. D 110~(9) (2024) 096018.
\newblock \href {http://arxiv.org/abs/2408.03106} {\path{arXiv:2408.03106}}, \href {https://doi.org/10.1103/PhysRevD.110.096018} {\path{doi:10.1103/PhysRevD.110.096018}}.

\bibitem{Florkowski:2024cif}
W.~Florkowski, {Spin hydrodynamics}, J. Subatomic Part. Cosmol. 3 (2025) 100028.
\newblock \href {http://arxiv.org/abs/2411.19673} {\path{arXiv:2411.19673}}, \href {https://doi.org/10.1016/j.jspc.2025.100028} {\path{doi:10.1016/j.jspc.2025.100028}}.

\bibitem{Bhadury:2025boe}
S.~Bhadury, Z.~Drogosz, W.~Florkowski, S.~K. Kar, V.~Mykhaylova, {Local equilibrium Wigner function for spin-1/2 particles} (5 2025).
\newblock \href {http://arxiv.org/abs/2505.02657} {\path{arXiv:2505.02657}}.

\bibitem{Drogosz:2025ihp}
Z.~Drogosz, W.~Florkowski, V.~Mykhaylova, {Application range of perfect spin hydrodynamics} (6 2025).
\newblock \href {http://arxiv.org/abs/2506.01537} {\path{arXiv:2506.01537}}.

\bibitem{Coleman:2018mew}
S.~Coleman, {Lectures of Sidney Coleman on Quantum Field Theory}, WSP, Hackensack, 2018.
\newblock \href {https://doi.org/10.1142/9371} {\path{doi:10.1142/9371}}.

\bibitem{Kapusta:2019sad}
J.~I. Kapusta, E.~Rrapaj, S.~Rudaz, {Relaxation Time for Strange Quark Spin in Rotating Quark-Gluon Plasma}, Phys. Rev. C 101~(2) (2020) 024907.
\newblock \href {http://arxiv.org/abs/1907.10750} {\path{arXiv:1907.10750}}, \href {https://doi.org/10.1103/PhysRevC.101.024907} {\path{doi:10.1103/PhysRevC.101.024907}}.

\bibitem{Kapusta:2020npk}
J.~I. Kapusta, E.~Rrapaj, S.~Rudaz, {Spin versus Helicity Equilibration Times and Lagrangian for Strange Quarks in Rotating Quark-Gluon Plasma}, Phys. Rev. C 102~(6) (2020) 064911.
\newblock \href {http://arxiv.org/abs/2004.14807} {\path{arXiv:2004.14807}}, \href {https://doi.org/10.1103/PhysRevC.102.064911} {\path{doi:10.1103/PhysRevC.102.064911}}.

\bibitem{deGroot:1980dk}
S.~R. De~Groot, {Relativistic Kinetic Theory. Principles and Applications}, Amsterdam, Netherlands: North-Holland 417p, 1980.

\bibitem{Frenkel:1926zz}
J.~Frenkel, {Die Elektrodynamik des rotierenden Elektrons}, Z. Phys. 37 (1926) 243--262.
\newblock \href {https://doi.org/10.1007/BF01397099} {\path{doi:10.1007/BF01397099}}.

\bibitem{Berestetskii:1982qgu}
V.~B. Berestetskii, E.~M. Lifshitz, L.~P. Pitaevskii, {Quantum Electrodynamics}, Vol.~4 of Course of Theoretical Physics, Pergamon Press, Oxford, 1982.

\bibitem{Florkowski:2019gio}
W.~Florkowski, A.~Kumar, R.~Ryblewski, {Spin Potential for Relativistic Particles with Spin 1/2}, Acta Phys. Polon. B 51 (2020) 945--959.
\newblock \href {http://arxiv.org/abs/1907.09835} {\path{arXiv:1907.09835}}, \href {https://doi.org/10.5506/APhysPolB.51.945} {\path{doi:10.5506/APhysPolB.51.945}}.

\bibitem{Itzykson:1980rh}
C.~Itzykson, J.~B. Zuber, \href{http://dx.doi.org/10.1063/1.2916419}{{Quantum Field Theory}}, International Series In Pure and Applied Physics, McGraw-Hill, New York, 1980.
\newline\urlprefix\url{http://dx.doi.org/10.1063/1.2916419}

\bibitem{Drogosz:2024lkx}
Z.~Drogosz, W.~Florkowski, N.~\L{}ygan, R.~Ryblewski, {Boost-invariant spin hydrodynamics with spin feedback effects}, Phys. Rev. C 111~(2) (2025) 024909.
\newblock \href {http://arxiv.org/abs/2411.06154} {\path{arXiv:2411.06154}}, \href {https://doi.org/10.1103/PhysRevC.111.024909} {\path{doi:10.1103/PhysRevC.111.024909}}.

\bibitem{Singh:2024cub}
S.~K. Singh, R.~Ryblewski, W.~Florkowski, {Spin dynamics with realistic hydrodynamic background for relativistic heavy-ion collisions}, Phys. Rev. C 111~(2) (2025) 024907.
\newblock \href {http://arxiv.org/abs/2411.08223} {\path{arXiv:2411.08223}}, \href {https://doi.org/10.1103/PhysRevC.111.024907} {\path{doi:10.1103/PhysRevC.111.024907}}.

\bibitem{Abboud:2025qtg}
N.~Abboud, L.~Gavassino, R.~Singh, E.~Speranza, {The perfect spinfluid} (6 2025).
\newblock \href {http://arxiv.org/abs/2506.19786} {\path{arXiv:2506.19786}}.

\bibitem{Becattini:2023ouz}
F.~Becattini, A.~Daher, X.-L. Sheng, {Entropy current and entropy production in relativistic spin hydrodynamics}, Phys. Lett. B 850 (2024) 138533.
\newblock \href {http://arxiv.org/abs/2309.05789} {\path{arXiv:2309.05789}}, \href {https://doi.org/10.1016/j.physletb.2024.138533} {\path{doi:10.1016/j.physletb.2024.138533}}.

\bibitem{Tolman:1934}
R.~C. Tolman, {Relativity, Thermodynamics and Cosmology}, Oxford University Press, London, 1934.

\bibitem{Klein:1949}
O.~Klein, {On the thermodynamical equilibrium of fluids in gravitational fields}, Rev. Mod. Phys. 21 (1949) 531.

\end{thebibliography}

\end{document}